# Some generalizations of the convective model of jet generation

S. N. Artekha[1,*]

[1]ORCID: 0000-0002-4714-4277

[1]Space Research Institute (IKI RAS), Moscow, Russia

[*]Corresponding author (sergey.arteha[at]gmail.com)

**Abstract.** For analytical description of the initial stage of jet generation in nonequilibrium inhomogeneous plasma in the magnetohydrodynamic approximation, possible generalizations of solutions of the nonlinear equation for the stream function are analyzed. The jet generation model is based on the mechanism of convective instability and the frozen-in condition of magnetic field lines and is characterized by a number of free parameters. The equation for the radial part of the stream function is satisfied by first-order Bessel functions. To satisfy all the conditions near the jet axis and on its periphery, the found solutions are smoothly joined at the boundary. The final analytical solution for the velocity field is applicable to arbitrary values of dimensionless coordinates. The poloidal velocity increases approximately exponentially, and the azimuthal velocity – according to a superexponential law. In this paper, the velocity field of the jet, which consists of seven sections, is calculated. The rotation of the jet turns out to be differential, and to obtain a solution in quadratures for the azimuthal velocity, one can use not only linear but also power dependences on the altitude. For the exponent n < 1, a noticeable increase in the azimuthal velocity with radius is observed immediately from the jet axis, and for n > 1, a region of relative calm is observed near the axis. The jet model is generalized to the case of an arbitrary dependence of the Brunt–Väisälä frequency on the altitude. The corresponding solutions are found for the radial and vertical velocity components. For the initial stage of development, the vertical and azimuthal components of the generated magnetic field of the jet were also found in the work.

**Key words:** inhomogeneous plasma, magnetosphere, jet, MHD approximation, Brunt–Väisälä frequency.

## 1. Introduction

Among the structures in magnetospheric, laboratory and space plasma, directed flows (jets) are quite often observed [1,2]. Such localized non-stationary objects are elongated vortex structures in which spiral motion is observed. In this case, the velocity of the plasma ejecta and the generated magnetic field reach maximum values in a certain region and decrease at the periphery of the structure. A wide class of jets observed in nature includes astrophysical jets [3,4,5], solar coronal jets and loops [6,7,8] and magnetospheric jets [9,10,11,12]. The study of dynamic structures arising in plasma is one of the fundamental problems of physics and is of great theoretical and practical interest. Numerous works are devoted to the development of the theory of plasma jets [13,14,15,16], the study of their dynamics, including taking into account the cyclotron motion of matter, [17,18,19,20]. Due to obvious limitations of existing models (see, for example, [1,2,4,20]), finding new solutions is an actual task. The creation of jet generation models opens the simplest and most correct way to obtaining a number of theoretically and practically interesting results. In [21], a low-parameter magnetohydrodynamic model for describing the initial stage of nonrelativistic jet development was constructed, based on the convective instability mechanism. The new analytical model allows one to describe the structure of the magnetic field and the velocity of a jet localized in space in the polar regions for all coordinate values and is an exact solution in the form of combinations of Bessel functions.

The aim of this paper is to generalize the proposed analytical model to the case of inhomogeneous plasma.

The structure of the paper is as follows. The Section 2 briefly derives the main equation for the stream function and the solutions that describe the developing instability. The Section 3 is devoted to generalizing the solutions of the obtained equation for an unstably stratified inhomogeneous plasma and determining the generated magnetic field. The Section 4 contains conclusions.

## 2. Basic equation for the current function

In the work [21], in the cylindrical coordinate system $(r,\varphi,z)$ for the axisymmetric case $\partial/\partial\varphi = 0$, the system of equations of ideal magnetohydrodynamics (MHD) was written out and weak perturbations of pressure, density and magnetic field were considered:

$$p = p_0(z) + \tilde{p}(t,r,z), \quad \rho = \rho_0(z) + \tilde{\rho}(t,r,z), \quad \mathbf{B} = \mathbf{B}_0 + \tilde{\mathbf{B}}, \tag{1}$$

where $\mathbf{B}_0 = (0,0,B_z)$; quantities $p_0$, $\rho_0$, $\mathbf{B}_0$ are equilibrium unperturbed values, and $\tilde{p}$, $\tilde{\rho}$ and $\tilde{\mathbf{B}}$ are small perturbations of the corresponding quantities at the initial stage of instability development. In the equilibrium state

$$\frac{d p_0}{d z} = -\rho_0(z)g, \quad \frac{d}{d r}\left(p_0 + \frac{B_0^2}{2\mu_0}\right) = 0, \tag{2}$$

where $\mu_0$ is the magnetic permeability of vacuum, $\mathbf{g} = -g\mathbf{e}_z$ is the gravitational acceleration on the Sun (star, or in the magnetosphere of the planet), $\mathbf{e}_z$ is a unit vector along the vertical. In the case of a divergence-free flow $\mathbf{v} = (v_r, v_\varphi, v_z)$, it is possible to introduce the stream function $\psi(t,r,\varphi,z)$:

$$v_r = -\frac{1}{r}\frac{\partial \psi}{\partial z}, \quad v_z = \frac{1}{r}\frac{\partial \psi}{\partial r}, \tag{3}$$

for which in [21, 22] from the system of MHD equations the following equation was obtained:

$$\left(\frac{\partial^2}{\partial t^2} + \omega_g^2\right)\Delta^*\psi + \frac{1}{r}\frac{\partial}{\partial t}J(\psi, \Delta^*\psi) = 0. \tag{4}$$

In this equation, the notations for the Jacobian are introduced

$$J(a,b) = \frac{\partial a}{\partial r}\frac{\partial b}{\partial z} - \frac{\partial a}{\partial z}\frac{\partial b}{\partial r}, \tag{5}$$

for the Grad–Shafranov operator (in the approximation of elongated structures $\partial/\partial r \gg \partial/\partial z$):

$$\Delta^* = r\frac{\partial}{\partial r}\left(\frac{1}{r}\frac{\partial}{\partial r}\right) \tag{6}$$

and for the square of the Brunt–Väisälä frequency:

$$\omega_g^2 = g\left(\frac{\gamma_a - 1}{\gamma_a H} + \frac{1}{T}\frac{dT}{dz}\right), \tag{7}$$

where $\gamma_a$ is the adiabatic index, $T$ is the gas temperature; is the characteristic height scale $H = \gamma p_0/g\rho_0$. Equation (4) is applicable to polar jets, when for the resulting magnetic field $\frac{L}{B_z}\frac{dB_z}{dz} \ll 1$. In particular, the linear case is considered in [21], where the addition to the poloidal vorticity from the emerging magnetic field is $-\frac{1}{\mu_0}\frac{\partial}{\partial r}\left(\frac{B_{0z}}{\rho}\frac{\partial \tilde{B}_z}{\partial z}\right)\mathbf{i}_\varphi \equiv 0$. In these cases, a purely convective mechanism operates.

In the case of $\omega_g^2 > 0$ equation (4) describes internal gravitational waves. We consider the opposite situation, when instability occurs at the moment $t = 0$, i.e. in (4) we have $\omega_g^2 \to -\gamma^2$. In this case, equation (4) describes localized dynamic structures growing with time. Such a situation arises if the internal regions of the plasma (the Sun, a star, or a magnetosphere) are hotter than the higher layers of the plasma: if the vertical temperature gradient (the second term in (7)) is negative, and its value exceeds the first term.

The solution in [21] was sought by the method of separation of variables:

$$\psi(t, r, z) = v_0 r^2 f(Z)\sinh(\gamma t)\Psi(R), \tag{8}$$

where $Z = z/L$, $R = r/r_0$, and $v_0$, $r_0$, $L$ are some characteristic speed, radial and vertical spatial scales. For the inner and outer regions, respectively, the following solutions were chosen:

$$\Psi_{int}(R) = \frac{J_1(\delta_0 R)}{R J_1(\delta_0)}, \tag{9}$$

$$\Psi_{ext}(R) = m \frac{K_1(\delta R)}{R K_1(\delta)}, \qquad (10)$$

where $\delta_0 \approx 1.841184$, parameter $\delta$ is arbitrary, and the value of $m$ is determined from the condition of smooth joining of the function and its derivatives at the boundary of the inner and outer regions.

To determine the azimuthal velocity in [21], the equation was used:

$$\frac{\partial v_\varphi}{\partial t} + \frac{v_r}{r}\frac{\partial}{\partial r}(rv_\varphi) + v_z \frac{\partial v_\varphi}{\partial z} = 0. \qquad (11)$$

It allows a solution by the method of separation of variables

$$v_\varphi = v_{\varphi 0} y(t) f_0(Z) V_{\varphi r}(R), \qquad (12)$$

if we choose a linear function $f(Z)$. In such a case, we obtain

$$y(t) = \exp\{c_0(\cosh(\gamma t) - 1)\}, \qquad (13)$$

and the coordinate dependence is determined by the equation

$$f'(Z)\left(\frac{\widehat{V}_r(R)}{R} + \frac{\widehat{V}_r(R)}{V_{\varphi r}(R)}\frac{dV_{\varphi r}(R)}{dR} + \frac{f(Z)f_0'(Z)\widehat{V}_z(R)}{f'(Z)f_0(Z)}\right) = -\frac{\gamma c_0 L}{v_0}. \qquad (14)$$

In [21], a linear function $f_0(Z) = f(Z)$ with a maximum at $Z = L/2$ was chosen. Then the radial dependence is expressed in quadratures. The corresponding graphs obtained for the velocity field components can be found in [21]. For ease of comparison, we will use the same free parameters of the model in this article.

### 3. Some generalizations of the jet model

To make the separation of variables method work, you can choose the following more general linear expression for the function

$$f(Z) = \begin{bmatrix} k_1 Z, & 0 \leq Z \leq Z_1; \\ \dfrac{k_1 Z_1}{1 - Z_1}(1 - Z), & Z_1 < Z \leq 1, \end{bmatrix} \qquad (15)$$

where $Z_1$ determines the coordinate of the maximum, and $k_1$ determines the rate of change with height. In this case, the most general expression for the function $f_0(Z)$ that allows a radial solution in quadratures will be the following:

$$f_0(Z) = \begin{bmatrix} k_2 Z^n, & 0 \leq Z \leq Z_1, \\ \dfrac{k_2 Z_1^n}{(1 - Z_1)^n}(1 - Z)^n, & Z_1 < Z \leq 1, \end{bmatrix} \qquad (16)$$

with an arbitrary exponent $n$. As a result, different differential rotations will be observed. The azimuthal velocity is determined by quadratures:

$$V_{\varphi r}^{int}(R) = v_{0\varphi} \exp\left\{\int_1^R \frac{-\alpha x - nx\widehat{V}_z^{int}(x) - \widehat{V}_r^{int}(x)}{x\widehat{V}_r^{int}(x)} dx\right\}, \quad (17)$$

$$V_{\varphi r}^{ext}(R) = Cv_{0\varphi} \exp\left\{\int_1^R \frac{-\alpha x - nx\widehat{V}_z^{ext}(x) - \widehat{V}_r^{ext}(x)}{x\widehat{V}_r^{ext}(x)} dx\right\}. \quad (18)$$

The obtained radial dependence can be easily calculated numerically (we took $\alpha = 0.01$) and depicted graphically using the Wolfram Mathematica program (Plot function). For $n < 1$, the rapid increase in the azimuthal velocity along the radius begins immediately from the jet axis, as can be seen from Fig. 1, and for $n > 1$, we see the rotation of a hollow cylinder (see Fig. 2), i.e., a region of relative calm is observed near the axis.

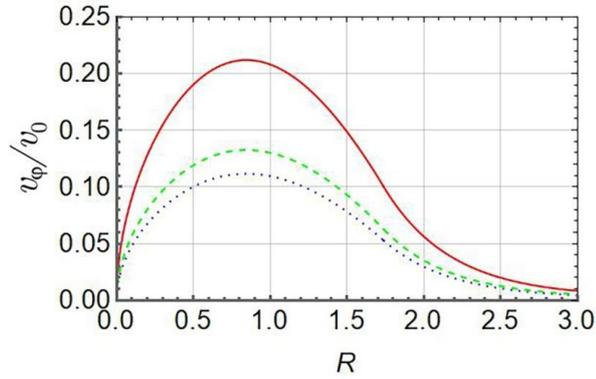

**Fig. 1.** Radial dependence of the relative azimuthal velocity $v_\varphi/v_0$ for $z/L=0.1$ and $n = 0.75$. The dotted line corresponds to the time $\gamma t = 3$, the dashed line corresponds to the time $\gamma t = 4$, and the solid line refers to the time $\gamma t = 5$

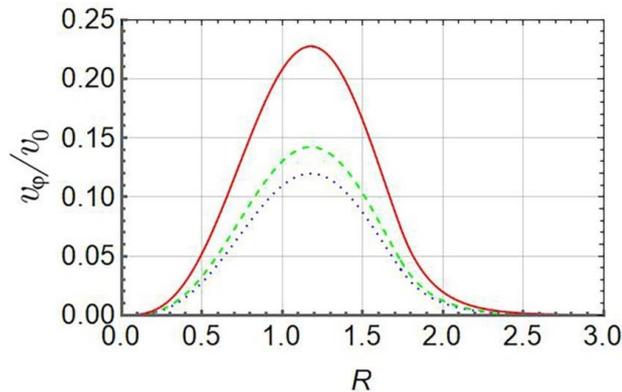

**Fig. 2.** Radial dependence of the relative azimuthal velocity $v_\varphi/v_0$ at $z/L=0.1$ and $n = 2$. The dotted line corresponds to the time $\gamma t = 3$, he dashed line corresponds to the time $\gamma t = 4$, and the solid line refers to the time $\gamma t = 5$

Note that the jet can be smoothly connected in height from more than two sections with different dependencies, like a constructor. For example, the velocity field $v_r$, $v_z$, $v_\varphi$ in Fig. 3 was calculated using the formulas obtained above for a jet consisting of seven sections along the vertical Z: (0,0.05); (0.05,0.1); (0.1,0.2); (0.2,0.5); (0.5,0.7); (0.7,0.9); (0.9,1). The graphic image was obtained using the Wolfram Mathematica program (DensityPlot function).

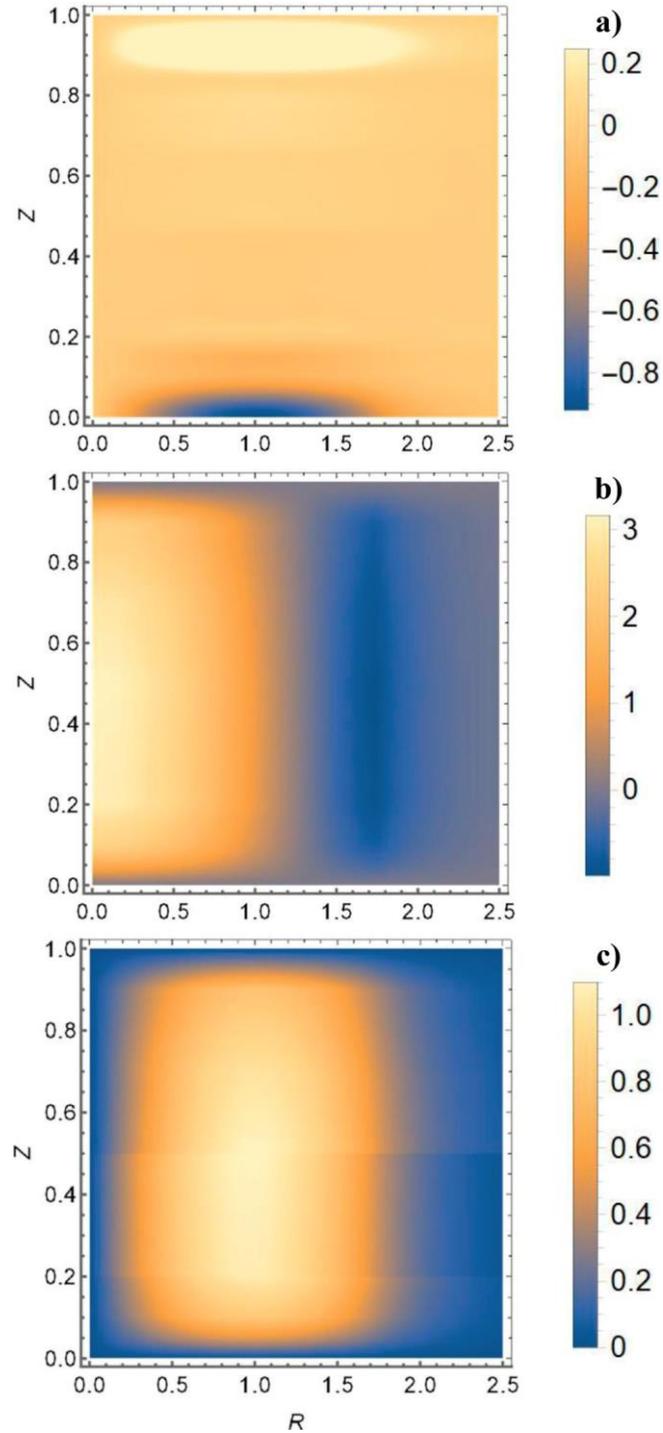

**Fig. 3.** The velocity field of a jet, which is constructed in height using seven linear sections of the function $f(Z)$, depending on the dimensionless coordinates $R$ and $Z$: a) $v_r(R,Z)$; b) $v_z(R,Z)$; c) $v_\varphi(R,Z)$.

In the solution for the poloidal velocity components, the function *f(Z)* can be chosen to be completely arbitrary, not just linear. However, in this case, as well as in the case of an arbitrary function $f_0(Z)$, the variables *R* and *Z* in equation (14) are not separated. Therefore, this equation must be solved simultaneously for two variables. Unfortunately, in many cases the numerical solution turns out to be unstable and quickly becomes turbulent. The jet is divided into many regions with differently directed azimuthal velocities, as can be seen from Fig. 4. The graphical representation was obtained using the Wolfram Mathematica program (DensityPlot function).

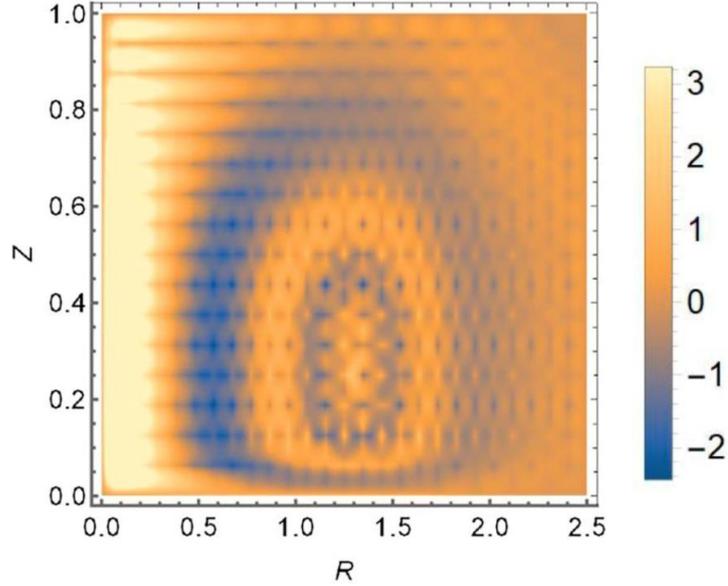

**Fig. 4.** An example of a multidirectional azimuthal jet velocity $v_\varphi(R,Z)$, depending on the dimensionless coordinates *R* and *Z*, for the case of a nonlinear dependence *f(Z)*.

Most likely, in such cases the solution is that not only differential rotation will be observed, but the rate of growth of the azimuthal velocity itself will also depend on the altitude, i.e. it is necessary to solve equation (11) for all three variables. Taking into account viscosity can also lead to a preferential direction of rotation of such turbulent plasma.

For a real system, the Brent-Väisälä frequency always depends on the height. Note the following important point related to the structure of equation (4). For an arbitrary *z*-dependence of the Brent-Väisälä frequency $\omega_g(z)$, the same stream function $\psi$ will remain the solution. The point is that the last term in (4) is satisfied only due to the radial dependence, and the first term is satisfied only due to the time dependence. Thus, in the expression for the stream function, there will be a single replacement: $\gamma \rightarrow \gamma(Z)$, i.e. the growth rate of all quantities will be differential with respect to *z*. As a result, in the expression for the vertical velocity component, there will be a single replacement: $\gamma \rightarrow \gamma(Z)$. However, in the expression for the radial velocity component, in addition to the replacement $\gamma \rightarrow \gamma(Z)$, the following replacement will also occur:

$f'(Z) \to f'(Z) + f(Z)\gamma'(Z)t \operatorname{Coth}[\gamma(Z)t]$. As a consequence, equation (11) can no longer be solved by the method of separation of variables, and the dependence on all three variables must be sought numerically.

The generated magnetic field can be found from the following equation (conditions for the magnetic field to be frozen into the plasma):

$$\frac{\partial \mathbf{B}}{\partial t} = \nabla \times (\mathbf{v} \times \mathbf{B}). \tag{19}$$

At the initial stage of generation, the final field can be found by the method of successive approximations:

$$B_z = B_0 + B_{z1} + B_{z2} + \cdots, \quad B_0 > B_{z1} > B_{z2},$$
$$B_\varphi = B_{\varphi 1} + B_{\varphi 2} + \cdots, \quad B_0 \gg B_{\varphi 1} > B_{\varphi 2}, \tag{20}$$
$$B_r = B_{r2} + \cdots, \quad B_0 \gg B_{r2},$$

where the equations for the corresponding corrections are as follows:

$$\frac{\partial B_{z1}}{\partial t} = B_0 \frac{\partial v_z}{\partial z}, \quad \frac{\partial B_{z2}}{\partial t} = B_{z1} \frac{\partial v_z}{\partial z} - v_r \frac{\partial B_{z1}}{\partial r}, \tag{21}$$

$$\frac{\partial B_{\varphi 1}}{\partial t} = B_0 \frac{\partial v_\varphi}{\partial z}, \quad \frac{\partial B_{\varphi 2}}{\partial t} = B_{z1} \frac{\partial v_\varphi}{\partial z} - v_r \frac{\partial B_{\varphi 1}}{\partial r}, \tag{22}$$

$$\frac{\partial B_{r2}}{\partial t} = B_{r2} \frac{\partial v_r}{\partial r} - v_r \frac{\partial B_{r2}}{\partial r} - v_z \frac{\partial B_{r2}}{\partial z}, \tag{23}$$

in doing so, starting with the second amendments, only numerical calculation is possible, but for the first amendments, solutions can be written in quadratures:

$$\tilde{B}_{z1}^{int} = \frac{B_0 v_0}{\gamma L} f'(Z)(\cosh(\gamma t) - 1) \delta_0 \frac{J_0(\delta_0 R)}{J_1(\delta_0)}, \tag{24}$$

$$\tilde{B}_{z1}^{ext} = -\frac{B_0 v_0}{\gamma L} f'(Z)(\cosh(\gamma t) - 1) m \delta \frac{K_0(\delta R)}{K_1(\delta)}, \tag{25}$$

$$\tilde{B}_{\varphi 1}^{int} = \frac{B_0 v_{\varphi 0}}{L} f'(Z) \exp\left\{ -\int_R^1 \frac{\alpha_{1,2} + V_z^{int}(x) - V_r^{int}(x)}{RV_r^{int}(x)} dx \right\} \int_0^t \exp\{c_0(\cosh(\gamma \tau) - 1)\} d\tau, \tag{26}$$

$$\tilde{B}_{\varphi 1}^{ext} = \frac{B_0 v_{\varphi 0}}{L} f'(Z) C \exp\left\{ \int_1^R \frac{\alpha_{1,2} + V_z^{ext}(x) - V_r^{ext}(x)}{RV_r^{ext}(x)} dx \right\} \int_0^t \exp\{c_0(\cosh(\gamma \tau) - 1)\} d\tau. \tag{27}$$

The generated vertical component of the magnetic field begins to grow exponentially in modulus (then the growth will gradually become superexponential). The azimuthal component of the magnetic field grows according to a superexponential law, but initially more slowly than the vertical component. The radial component of the magnetic field grows even more slowly and always remains the smallest of these three field components. For the case of seven linear sections of the function *f(Z)*, the distribution of the resulting vertical component and the azimuthal component of the magnetic field are shown in Fig. 5 a) and b), respectively. The graphical representation was obtained using the Wolfram Mathematica program (DensityPlot function).

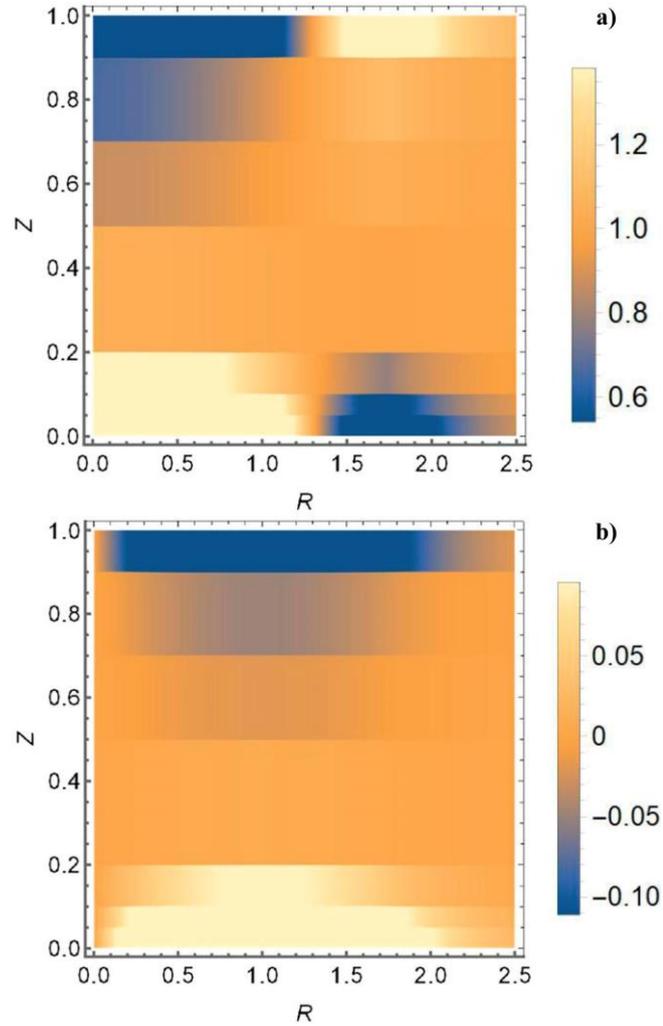

**Fig. 5.** The resulting dimensionless magnetic field at the initial stage of growth as a function of the dimensionless coordinates *R* and *Z*: a) $B_z(R,Z)/B_0$; b) $B_\varphi(R,Z)/B_0$. The jet is constructed in height using seven linear sections of the function *f(Z)*.

After some time, the generated non-uniform magnetic field can exceed the initial uniform field. Of course, at later times, when moving to the quasi-stationary stage, it is necessary to take into account dissipative processes and the influence of the generated field on the jet generation process.

# 4. Conclusion

In this paper, within the framework of ideal MHD, a nonlinear equation is presented for the stream function, which in unstable stratified plasma describes the formation of axially symmetric structures growing in time. The resulting equation can be solved by separation of variables, which allows us to obtain first-order Bessel functions as solutions for the radial part. To satisfy all the conditions near the jet axis and on its periphery, the found solutions are smoothly joined in the intermediate region. As a result, an analytical solution is obtained for the velocity field applicable for all values of the dimensionless coordinates $R$ and $Z$. The jet model is characterized by a number of free parameters: the characteristic radial velocity $v_0$ and the initial azimuthal velocity $v_{0\varphi}$; the characteristic radial scale of the jet $r_0$ and the vertical scale $L$; the parameter $m$, which characterizes the radial structure of the jet; a uniform external magnetic field $B_0$ and the increment of convective instability $\gamma$. The jet can be constructed vertically from a number of sections. In this paper, the velocity field is calculated for a jet, which is composed of seven linear sections. The poloidal velocity increases approximately exponentially, and the azimuthal velocity increases superexponentially. The jet rotation turns out to be differential, and to obtain a solution in quadratures for the azimuthal velocity, one can use not only linear but also power dependences on the height $Z^n$. For $n < 1$, the rapid increase in the azimuthal velocity along the radius begins immediately from the jet axis, and for $n > 1$, a region of relative calm is observed near the axis. The jet model is generalized to the case of an arbitrary $z$-dependence of the Brent – Väisälä frequency $\omega_g(z)$. The corresponding solutions for the radial and vertical velocity components are found. In the case of an arbitrary dependence on the height, the equation for the azimuthal velocity cannot be solved by separating the variables, and the dependence on all three variables must be sought numerically. In this case, solutions with multidirectional rotations can be obtained. For the initial stage of development, the vertical and azimuthal components of the generated magnetic field of the jet, which consists of seven linear sections, were also found in the work.

Thus, in convective-unstable plasma, intense jets are formed extremely quickly. The generalized model describes the generation of the velocity field and magnetic field, the magnitude of which increases with time.